\documentclass[pdflatex,sn-mathphys-ay]{sn-jnl}
\usepackage{graphicx}
\usepackage{multirow}
\usepackage{amsmath,amssymb,amsfonts}
\usepackage{amsthm}
\usepackage{mathrsfs}
\usepackage[title]{appendix}
\usepackage{xcolor}
\usepackage{textcomp}
\usepackage{manyfoot}
\usepackage{booktabs}
\usepackage{algorithm}
\usepackage{algorithmicx}
\usepackage{algpseudocode}
\usepackage{listings}
\usepackage{url}
\usepackage{float}
\begin{document}

\title[Judgement in the Age of Jev]{Judgement in the Age of Jev: From Evaluation Scarcity to Evaluation Abundance}

\author{\fnm{Richard} \sur{Hill}}
\affil{\orgdiv{Department of Computer Science}, \orgname{University of Huddersfield}, \orgaddress{\city{Huddersfield}, \country{UK}}}
\affil{\normalfont\texttt{r.hill@hud.ac.uk}}
\affil{\normalfont ORCID: \href{https://orcid.org/0000-0003-0105-7730}{0000-0003-0105-7730}}

\abstract{Generative artificial intelligence has reduced the cost of producing plausible symbolic artefacts, leading recent organisation scholarship to identify evaluation and discernment as constraints under conditions of production abundance. This perspective examines a further possibility: that machine evaluation itself becomes inexpensive enough to be deployed routinely and at scale. The investigation is prompted by Jev, TypeSafe AI's specialised model for typed probabilistic decisions. TypeSafe explicitly invokes William Stanley Jevons to argue that lower-cost machine intelligence can unlock previously uneconomic uses. Treating this as a technological provocation rather than an established empirical result, the article formulates a conditional Jevons hypothesis for machine evaluation: sufficiently large reductions in the total marginal cost of usable machine evaluation may increase its organisational consumption where latent demand is substantial and complementary costs do not dominate. The article integrates rebound economics with research on cheap prediction, production abundance, machine evaluation, decision allocation, authority, reliance and Executive Judgement to examine this possible scarcity transition. It distinguishes prediction, machine evaluation, organisational judgement and authorisation as functional activities whose costs need not fall together. This distinction clarifies why greater evaluative throughput need not yield proportionately greater organisational warrant. Evaluations can share evidence, criteria and errors; scale misspecified rubrics; operate on representations from which consequential qualifications have disappeared; and change practical decision rights through thresholds and exception routing. The resulting research problem is when cheap machine evaluation substitutes for human evaluative work, when it redistributes or creates demands for judgement, and how it affects the grounds available at consequential organisational commitment. The contribution is a focused scholarly synthesis and research agenda for a possible transition from evaluation scarcity towards evaluation abundance.}

\keywords{Jev, machine evaluation, organisational judgement, executive judgement, Jevons paradox, evaluation abundance, human--AI decision-making, epistemic boundedness.}

\maketitle

\section{Introduction}

Generative artificial intelligence has altered the economics of producing organisational artefacts. Memos, analyses, code, proposals, plans and alternative formulations that once consumed substantial human time can now be produced quickly and at low marginal cost. Productivity studies already show material gains in bounded professional tasks, while organisation scholarship increasingly asks what happens when the rate of plausible production exceeds the capacity available to evaluate and integrate it (Noy and Zhang, 2023; Brynjolfsson et al., 2025; Cao et al., 2026). Cao et al. (2026) describe this condition as a \emph{discernment gap}: production becomes abundant while coherent evaluation, prioritisation and learning remain comparatively scarce.

If generative AI has made production abundant, an obvious next question follows: what happens when evaluation itself becomes cheap? The immediate provocation is Jev, the first public ``System One Model'' released by TypeSafe AI. TypeSafe describes Jev as a model for fast, structured decisions that software can consume directly: unstructured state is converted into typed probabilistic decisions rather than generated prose. The company presents classification, scoring, routing, judging, verification and guardrailing as candidate uses and reports large cost and latency advantages over its selected large-language-model workflows (Almeida, 2026). Those performance figures are vendor-reported, workload-dependent and not treated here as independent scientific evidence. The more important point is architectural and economic. Jev is an explicit attempt to make bounded machine decisions sufficiently inexpensive and fast to be embedded pervasively in software workflows.

The name is consequential to the argument. TypeSafe states that Jev is named after William Stanley Jevons and explicitly invokes the logic associated with the Jevons paradox: improvements in efficiency can increase aggregate consumption because lower effective cost makes additional uses viable (Almeida, 2026). The company's proposition is that large reductions in the cost of machine intelligence will unlock many more uses. This article asks a narrower organisational question. If the marginal cost of machine \emph{evaluation} falls sufficiently, might organisations consume much more evaluation rather than merely perform their existing evaluations more cheaply? If so, what follows for organisational and executive judgement?

The question extends, but does not reject, the economics of cheap prediction. Agrawal et al. (2019) conceptualise contemporary machine learning as a prediction technology and distinguish prediction from the judgement and action with which it is combined in decisions. Their analysis shows why a reduction in the cost of one cognitive input can change the value and organisation of complementary activities. Jev-like systems suggest that a related economic change may be reaching evaluative tasks themselves. A machine can increasingly be asked not only what is likely to happen, but whether an artefact satisfies a criterion, which of several alternatives ranks highest, whether a case should be escalated, or whether an output passes a specified check. These are bounded evaluative acts. They need not constitute organisational judgement in the fuller sense of deciding what grounds matter and whether those grounds warrant consequential commitment.

That distinction is important because evaluation abundance does not imply evidential or judgement abundance. Research on algorithmic reliance has long shown that human use of machine advice is conditional and can exhibit both aversion and appreciation (Dietvorst et al., 2015; Logg et al., 2019). Research on automation distinguishes use, misuse and disuse and emphasises appropriate rather than maximal reliance (Parasuraman and Riley, 1997; Lee and See, 2004). More recent work on generative AI identifies an additional epistemic problem: abundant outputs can be difficult to verify because their veracity and foundations are not necessarily transparent (Kim et al., 2026). Machine evaluators add a further complication. Multiple evaluations may be correlated rather than independent; a recent study of nine frontier LLM judges found that their joint information was approximately equivalent to only two independent votes on the tested tasks (Kohli, 2026). More evaluation therefore need not mean proportionately more independent grounds.

The tempting conclusion is that cheaper evaluation simply removes the next bottleneck. The argument here is that the answer is not so simple. Prediction, evaluation, organisational judgement and authorisation perform different functions, and their costs need not fall together. The article therefore proposes a conditional \emph{Jevons hypothesis for machine evaluation}: sufficiently large reductions in marginal evaluation cost may increase the quantity, frequency, granularity and scope of organisational evaluation where latent demand is substantial and complementary costs do not dominate. The organisational question is what happens to judgement when that occurs.

This question connects directly to Hill (2026), \emph{Executive Judgement in AI-Mediated Decision-Making Environments}. That work argues that locally competent contributions can still culminate in inadequately warranted organisational commitment when consequential assumptions, uncertainty, dependencies or dissent are weakened during transformation before authorisation. Cheap evaluation does not displace that problem. It may create many more occasions on which evaluation enters the formation process, while leaving unresolved whether the evidence, criteria and qualifications available at authorisation are adequate. The present article therefore treats Executive Judgement as a theoretical lens rather than claiming a second process theory.

Three research questions organise the analysis:

\begin{enumerate}
\item How might falling machine-evaluation costs change the quantity, frequency, granularity and organisational distribution of evaluation?
\item Under what conditions does abundant machine evaluation augment, substitute for or generate new demands for organisational judgement?
\item What are the consequences of evaluation abundance for the formation, challenge and authorisation of consequential organisational commitments?
\end{enumerate}

The contribution is a scholarly synthesis and research agenda. It distinguishes cheap prediction, cheap production and cheap evaluation; examines a plausible transition from evaluation scarcity to evaluation abundance; and connects that transition to the problem of warranted organisational commitment. The Jevons rebound remains a hypothesis to be tested, Jev's current performance remains vendor-reported and domain-dependent, and the consequences for executive judgement remain contingent on how work and decision rights are organised.

\section{Conceptual approach and scope}

This article uses an integrative conceptual synthesis rather than a systematic review or empirical study. The investigation began with a specific technological observation: a model explicitly designed to make structured machine decisions inexpensive enough for pervasive software use. The search then proceeded destructively, asking whether the proposed organisational consequences were already explained by established work on cheap prediction, human--AI decision structures, automation and augmentation, organisational evaluation, epistemic constraints, reliance, distributed cognition and authority.

The focal unit is a consequential organisational decision episode in which artefacts or cases are produced, evaluated, transformed and connected to action. \emph{Evaluation} refers to assessing an object, case, representation or output against an explicit or implicit criterion. It includes activities such as scoring, ranking, classification, critique, verification and selection. \emph{Organisational judgement} refers more broadly to determining which grounds matter in a situation, how competing grounds should be interpreted and what conclusion the organisation should draw from them. \emph{Authorisation} concerns whether those grounds are sufficient for a role or process with decision rights to commit organisational action or resources.

These are functional distinctions rather than claims about mutually exclusive mental processes. Prediction can be embedded inside evaluation; evaluation can influence judgement; and a policy can pre-authorise actions that remove a contemporaneous human authoriser. The decomposition is useful because technological change can alter the cost of one function without reducing the costs or difficulties of the others by the same amount.

TypeSafe's claims about Jev establish what the vendor is attempting to build and how it motivates the product. They do not establish a general rebound effect in organisational evaluation. The Jevons hypothesis is therefore conditional and prospective. Its value lies in asking what becomes theoretically and empirically important if machine evaluation becomes much cheaper, whether through Jev, future System One Models, specialised classifiers, LLM-based judges or another architecture.

\section{From cheap prediction to cheap evaluation}

\subsection{Prediction as an input to decisions}

Agrawal et al. (2019) argue that much of machine learning's economic significance can be understood as a reduction in the cost of prediction. Prediction is valuable because it is an input into decisions, but a decision also requires an action set and a judgement about payoffs. Their decomposition is important here because it resists treating all machine cognition as interchangeable. When prediction becomes cheaper, organisations may use more prediction, reorganise tasks around it and change the relative value of complementary judgement (Agrawal et al., 2018, 2019).

The contemporary generative-AI transition adds another change in relative cost. Generative systems reduce the effort required to produce plausible symbolic artefacts. Noy and Zhang (2023) demonstrate productivity effects in professional writing tasks, while Brynjolfsson et al. (2025) report productivity gains in customer-support work. Neither study establishes general organisational transformation, but together with the rapid diffusion of generative systems they motivate attention to the economics of production rather than prediction alone.

Cao et al. (2026) make this organisational consequence explicit. They argue that generative AI can make plausible artefact production abundant while evaluation practices remain adapted to a lower-volume regime. Episodic review, expert gatekeeping and ad hoc escalation become strained when production grows faster than the capacity to decide what is correct, distinctive, strategically relevant or worth pursuing. Their proposed response is evaluation-centred organisation design involving continuous curation, distributed calibration, evaluative memory and attention governance. The important premise for the present paper is that evaluation is treated as the comparatively scarce complement to abundant production.

\subsection{Jev as a technological provocation}

Jev potentially changes that premise. TypeSafe's public description positions the model as a machine-native decision primitive rather than a conversational generator. The model returns typed decisions with probabilities and confidence estimates intended to be consumed by software. TypeSafe further suggests that thresholds can determine when software acts autonomously and when it escalates for review (Almeida, 2026; TypeSafe AI, 2026). This is organisationally important because evaluation can become part of the control flow of work rather than a separate human review stage.

The economic claims should be separated from the architectural claim. TypeSafe currently reports Jev workflows that are substantially faster and cheaper than its LLM comparators and prices the service at a very low input-token rate (TypeSafe AI, 2026). These figures are not independent evidence that all evaluation will become two orders of magnitude cheaper. Workflows and external integration costs have impact, and a low inference price does not eliminate the costs of specifying criteria, preparing evidence, validating outputs, reviewing exceptions or correcting errors. The scholarly premise required here is weaker: specialised machine evaluation may reduce the marginal cost of at least some bounded evaluative acts sufficiently to change where organisations can economically deploy them.

This possibility is consistent with a wider technical movement towards machine evaluation. Li et al. (2025) define the LLM-as-a-judge paradigm around machine scoring, ranking and selection across evaluation settings. Evidence is also beginning to extend beyond evaluation of machine outputs. Addis et al. (2026), for example, compare repeated ChatGPT evaluations of human creative work with expert human judgements in constrained innovation-like tasks and find comparable internal consistency, while also reporting systematic differences in score level and dispersion. These results establish neither general evaluator competence nor organisational judgement, but they show that some evaluative work formerly requiring human panels can be machine-mediated at scale. Such systems should not be treated as interchangeable with Jev. Jev's non-generative, typed-output design is specifically intended for software automation. The common organisational feature is that some bounded evaluative acts that previously consumed human time can increasingly be delegated to machine processes at low marginal cost.

\section{Positioning against the closest scarcity accounts}

A claim that AI makes ``judgement'' abundant or scarce is too coarse to support the present argument. Several contemporary accounts already describe scarcity inversions around AI, but they use different units of analysis and different meanings of judgement. The contribution here depends on distinguishing them rather than claiming an empty field.

Cao et al. (2026) provide the closest organisational starting point. Their discernment-gap argument holds that generative AI lowers the cost of producing plausible symbolic artefacts faster than organisations can coherently evaluate, prioritise and learn from them. Evaluation-centred organisation design is proposed as a response to this imbalance. The present paper accepts rather than contests that diagnosis. Its question concerns a possible subsequent regime: what if some bounded evaluation also becomes sufficiently inexpensive and software-native to scale with production?

Lov\'en (2026) goes further and explicitly argues for a post-scarcity of competent-looking judgement, including selecting, ranking, attributing and certifying. He identifies verified signal, legitimacy, authentic provenance and integration capacity as emerging complements. This is close to the phenomenon considered here. The difference is analytical vocabulary and the organisational endpoint. The present paper calls these bounded outputs \emph{machine evaluations} so that their scalable production is not conflated with the organisational judgement required to determine which grounds are important and whether they warrant commitment. This is a functional distinction, not a claim that the term judgement can only be used in one way.

West and Cosimano (2026) offer a different scarcity account. Their Judgment Cost Theory defines judgement cost around accountable action under irreducible uncertainty and argues that such judgement becomes relatively scarce as prediction becomes cheap. Baaij (2026), meanwhile, describes analytical abundance in strategy: greater analytical capacity can sustain multiple credible futures, moving the strategic constraint from insight generation towards choice and commitment. Both accounts caution against the generic claim that more machine cognition simply makes ``judgement'' abundant. They also reinforce the need to specify which activity has become cheaper and which complement remains consequential.

Taken together, these accounts rule out several stronger claims. This article does not claim novelty for the idea that AI can invert cognitive scarcities, that evaluation can become a bottleneck, that abundant analysis can move constraints towards commitment, or that accountability can remain scarce. Its narrower synthesis concerns the transition in the \emph{organisational consumption of machine evaluation}. The question is whether a sufficiently large reduction in the cost of bounded evaluative acts changes their quantity, frequency, granularity and placement, and what established organisational theories imply when it does.

\begin{table}[htbp]
\caption{Adjacent scarcity accounts and the residual question}\label{tab:scarcity}
\centering
\small
\begin{tabular}{@{}p{0.22\textwidth}p{0.31\textwidth}p{0.39\textwidth}@{}}
\toprule
Account & Scarcity transition & Relevance to this article \\
\midrule
Agrawal et al. (2019) & Prediction becomes cheaper & Separates prediction from complementary judgement and action. \\
Cao et al. (2026) & Production becomes abundant while discernment remains scarce & Establishes evaluation scarcity as an organisational problem under generative abundance. \\
Lov\'en (2026) & Competent-looking judgement becomes abundant & Closest post-scarcity account; motivates a stricter distinction between machine evaluation and organisational warrant. \\
West and Cosimano (2026) & Cheap prediction makes accountable judgement relatively scarce & Shows that accountable commitment can remain costly even as machine cognition becomes cheap. \\
Baaij (2026) & Analytical abundance shifts constraint towards choice and commitment & Establishes that abundance can migrate a strategic constraint without implying a new causal mechanism. \\
Present synthesis & Bounded machine evaluation may become cheaper and more widely consumed & Asks when evaluation rebounds and how expanded evaluation changes the grounds and demands of organisational judgement. \\
\bottomrule
\end{tabular}
\end{table}

\section{A Jevons hypothesis for machine evaluation}

The Jevons paradox originates in the observation that efficiency improvements need not reduce aggregate resource consumption when lower effective cost stimulates additional demand. Rebound reasoning has also been imported into organisation and management scholarship, where Schaefer and Wickert (2015) use Jevons logic to question the assumption that efficiency gains translate straightforwardly into lower aggregate resource use. TypeSafe explicitly invokes this logic in naming Jev, stating that reductions in the cost of machine intelligence should unlock additional uses (Almeida, 2026). The analogy should not be converted into a law. Rebound depends on demand response, substitution, complementary costs and the boundaries of the activity being measured.

For organisational evaluation, the relevant hypothesis can be stated as follows:

\begin{quote}
\textbf{Jevons hypothesis for machine evaluation.} Where reductions in the marginal cost and latency of machine evaluation are sufficiently large, latent demand for evaluation is substantial, and complementary costs do not dominate, organisations will increase the aggregate consumption of evaluation by deploying evaluative acts that were previously uneconomic.
\end{quote}

This is a research hypothesis rather than a proposition claimed as established theory. It predicts more than cost saving. A firm that previously sampled ten per cent of transactions could evaluate all transactions. A proposal might be assessed against twenty criteria rather than three. Evaluation might move from final outputs to intermediate states, from periodic review to continuous checking, or from a single evaluator to multiple evaluators. Evaluation could itself become recursive, with one process checking the outputs, confidence or agreement of another.

The hypothesis also has clear failure conditions. Evaluation consumption may not rise materially if the organisation has little latent demand, if additional evaluations have low marginal value, if integration and review costs dominate inference cost, if legal or professional constraints limit automation, or if confidence in the evaluator is insufficient. The rebound could also be partial: routine evaluations might become abundant while high-stakes or context-rich evaluation remains scarce.

This conditional formulation is important because ``cheap intelligence'' can conceal heterogeneous activities. An inexpensive model call is not equivalent to an inexpensive organisational evaluation. Evidence must be assembled, criteria specified, outputs routed and exceptions handled. The relevant economic quantity is therefore the total cost of a usable evaluative act within a workflow, not inference price alone.

\section{Evaluation abundance is not judgement abundance}

\subsection{Evaluation competence and organisational warrant}

A central risk in discussing AI and judgement is to treat any machine output that contains an assessment as equivalent to organisational judgement. The distinction is clearer if the functions are decomposed. Prediction asks what is likely to be true or happen. Evaluation asks how an object or case performs against some criterion. Organisational judgement asks what grounds matter and what should be concluded from them. Authorisation asks whether those grounds are sufficient for consequential commitment.

Agrawal et al. (2019) already distinguish prediction from judgement. The present argument extends that decomposition to evaluation. A system can competently determine that a proposal satisfies a stated rubric without establishing that the rubric captures what the organisation ought to care about. It can verify that a document contains required fields without establishing that the underlying evidence is adequate. It can rank alternatives according to specified dimensions without determining whether a missing dimension is consequential.

Such a contrast is compatible with research on bounded rationality and distributed cognition. Simon's (1955) account of bounded rationality explains why decision processes operate under limits of information and computation. Hutchins (1995) shows that cognition can be distributed across people and artefacts rather than located solely in an individual mind. AI-mediated evaluation can relax some computational constraints and participate in distributed cognitive work without thereby eliminating the organisational problem of deciding which representations and grounds should govern commitment.

\subsection{Grounding and epistemic boundedness}

Kim et al. (2026) argue that generative AI changes administrative decision-making by creating a condition of \emph{epistemic boundedness}: information can be abundant while decision makers remain unable to verify its veracity and foundations. Their argument concerns generative AI and administrative rationality, not machine evaluation specifically. It nevertheless identifies a closely related constraint. An evaluative output is useful only to the extent that its evidential basis and relationship to the decision problem are adequate.

Machine evaluation can therefore increase evaluative throughput without increasing evidential sufficiency proportionately. If an evaluator receives a summary from which a material qualification has disappeared, perfect application of the evaluation criterion cannot restore information it never received. This is the connection to Executive Judgement. Qualification attrition concerns the weakening or loss of consequential assumptions, uncertainty, dependencies or dissent as representations are transformed before authorisation. Adding more evaluation downstream of the same lossy representation may multiply competent assessments without repairing the missing grounds.

\subsection{Independence and apparent corroboration}

Evaluation abundance also creates a problem of dependence. Organisational decision processes often treat agreement among evaluators as corroborating evidence. That inference is weaker when evaluators share training data, prompts, evidence, rubrics or systematic errors. Kohli (2026) provides a technical demonstration, showing that across three natural-language-inference datasets, nine `frontier' LLM judges from seven model families provided approximately two independent votes' worth of information; correlated errors substantially reduced the benefit expected under independent voting.

The ratio is specific to the tested models and tasks. The more general point is that the nominal number of machine evaluations cannot be treated as the effective amount of independent evidence. More judges can produce more outputs while adding relatively little new information. Subsequent work further suggests that aggregate dependence metrics and the local value of an independent verification signal are distinct questions, reinforcing the need to examine where independent evidence changes a decision rather than simply counting evaluators (Shu, 2026).

The organisational implication is not that evaluator panels are useless. It is that evaluation abundance changes the burden of understanding provenance and dependence. If organisations can cheaply create dozens of evaluators, they also need ways to determine whether those evaluators provide distinct grounds or merely reproduce the same evidential structure.

\subsection{Reliance and evaluative framing}

Human response to algorithmic advice is itself variable. Dietvorst et al. (2015) show algorithm aversion after observed error, while Logg et al. (2019) demonstrate algorithm appreciation in other experimental settings. Lee and See (2004) distinguish trust from appropriate reliance, and Parasuraman and Riley (1997) distinguish use, misuse and disuse. These findings caution against a simple assumption that abundant machine evaluation will either dominate human judgement or be routinely ignored.

A further possibility is that machine evaluation changes the frame within which human judgement is exercised. Once work arrives pre-scored, pre-ranked or pre-screened, the human task can shift from forming an assessment independently to interpreting or contesting an existing machine evaluation. This is not necessarily harmful. Pre-evaluation may focus scarce attention on anomalies and improve consistency. But it can also narrow the set of questions that receive attention if the machine's criteria become the de facto structure of review. The effect is therefore an empirical question about task design, reliance and authority rather than a universal consequence of automation.

\section{Where constraints may migrate}

If machine evaluation becomes abundant, the organisational bottleneck need not disappear. It may migrate towards complementary resources whose costs do not fall at the same rate. Five such constraints are particularly salient in the existing literature and the Executive Judgement framework.

\subsection{Grounding}

Evaluation requires an object and an evidential representation. The quality of an evaluator cannot compensate for material evidence that is absent, inaccessible or transformed beyond usefulness. Epistemic boundedness therefore remains relevant even when evaluative capacity is abundant (Kim et al., 2026). In Executive Judgement terms, qualification preservation becomes important because evaluation can only operate on the grounds that survive to it.

\subsection{Criteria}

Cheap application of criteria is different from cheap determination of appropriate criteria. A model can repeatedly ask whether an investment meets a hurdle rate, whether a contract contains specified clauses or whether a case crosses a risk threshold. The normative and strategic question of whether those are the right criteria can remain contested. Nonaka and Toyama's (2007) account of distributed practical wisdom is relevant because appropriate action depends upon particular circumstances and purposes, not merely the mechanical application of a general rule. Lindebaum et al. (2026) similarly examine managerial phronesis under generative AI, reinforcing the distinction between computational assistance and context-sensitive practical judgement.

\subsection{Independence}

Cheap replication can make apparent corroboration abundant. Yet correlated machine evaluators may add much less independent evidence than their number suggests (Kohli, 2026). Independence therefore becomes an architectural rather than numerical property. It may require different evidence sources, methods or human expertise rather than additional instances of the same evaluator family.

\subsection{Interpretive attention}

Cao et al. (2026) treat attention governance as central when production is abundant and evaluation remains scarce. Machine evaluation could reduce some of that scarcity, but it can also generate new material: scores, flags, disagreements, exceptions and confidence estimates. Human attention may consequently move rather than disappear. The relevant question becomes which machine-produced signals deserve interpretation and escalation.

\subsection{Authority and intervention}

Decision structures determine how machine outputs connect to action. Shrestha et al. (2019) distinguish forms of human--AI decision-making involving delegation, sequential decision processes and aggregation. Parasuraman et al. (2000) similarly show that automation can operate at different stages and levels of decision activity. Jev-like systems make this design choice especially concrete because confidence thresholds can be encoded to determine whether software acts, seeks another evaluation or escalates to a human (TypeSafe AI, 2026).

A threshold is therefore not merely a technical parameter. In an organisational workflow it can allocate attention and practical decision rights. Formal human accountability provides limited assurance if the accountable role lacks access to consequential grounds or effective ability to intervene. The distinction between formal and real authority is long established in organisation theory (Aghion and Tirole, 1997), while Elish (2019) shows how responsibility can be displaced onto human operators in complex automated systems. This is consistent with the Executive Judgement distinction between nominal approval and meaningful authorisation.

There is unlikely to be one resource that simply replaces evaluation as the new universal bottleneck. Different organisational settings will have different binding constraints. In routine, well-specified tasks, abundant evaluation may remove substantial human work. In ambiguous or high-stakes settings, criterion selection, evidence interpretation or authorisation may dominate. The useful theoretical move is therefore from a single-bottleneck account to a conditional question about where constraints migrate as evaluation cost falls.

\section{Consequences for organisational and executive judgement}

\subsection{Substitution, augmentation and new demand}

The automation--augmentation paradox already cautions against treating substitution and complementarity as mutually exclusive organisational futures (Raisch and Krakowski, 2021). Hybrid problem-solving research similarly examines combinations of human and artificial intelligence rather than a simple contest between them (Raisch and Fomina, 2025). Evaluation abundance should be analysed in the same way.

First, machine evaluation can substitute for human evaluative work where criteria are stable, evidence is accessible and errors are sufficiently measurable. Routine screening, routing and compliance checks are obvious candidates. Second, machine evaluation can augment human judgement by supplying additional checks, comparisons or anomaly detection. Third, evaluation abundance can create new judgement demand by surfacing exceptions, disagreements or criterion conflicts that were previously invisible because the underlying evaluation was never performed.

The third possibility is a deduction, not an established general empirical result. It follows from the possibility that cheaper evaluation increases the number of cases receiving scrutiny. If only a small proportion of a much larger evaluated population requires human interpretation, absolute human review demand could still rise. Conversely, improved automation could reduce it. The direction depends on evaluation volume, exception rates, workflow design and the cost of resolution.

\subsection{Pre-evaluated organisational work}

A further consequence concerns the composition of material reaching managers and executives. When evaluation is scarce, decision makers may encounter relatively unevaluated cases and perform substantial first-order assessment themselves. Under evaluation abundance, more cases may arrive already classified, scored, ranked, critiqued or filtered. Executive work can therefore become increasingly concerned with what to make of evaluations rather than producing every evaluation directly.

That shift need not move judgement upwards to executives. Shrestha et al. (2019) show that decision structures can allocate human and AI roles in multiple ways, while distributed cognition cautions against locating all relevant cognition at the top of a hierarchy (Hutchins, 1995). The stronger implication is simpler: abundant machine evaluation changes the material on which organisational judgement operates and can therefore change where interpretive and authorising work is located.

\subsection{Connection to Executive Judgement}

Executive Judgement provides a process lens for analysing this changed environment. Its central concern is the movement from distributed judgement formation to consequential authorisation and the possibility that material qualifications are lost during transformation. Evaluation abundance potentially affects each stage.

At formation, organisations can insert many more machine evaluations into a decision episode. This may expose unsupported assumptions, compare alternatives and detect anomalies. It may also standardise attention around machine-readable criteria. During transformation, machine-generated summaries and scores can preserve some information while compressing other information. At challenge, multiple evaluators can increase scrutiny but may provide less independent challenge than their number implies. At authorisation, confidence thresholds and routing policies can determine which cases remain visible to a human decision authority.

Jev does not make Executive Judgement more valuable by definition. It makes the relationship between evaluation and warranted commitment harder to ignore because evaluation can be produced at scale. A decision process can contain many competent evaluations while still presenting an authorising role with inadequate grounds if the evaluations share a missing premise, rely on inappropriate criteria or conceal consequential qualifications. Conversely, abundant evaluation could improve organisational warrant when it restores missing evidence, introduces genuinely independent challenge or directs attention towards consequential exceptions.

This yields a useful distinction between \emph{evaluation throughput} and \emph{warranted commitment}. Jev-like technologies directly target the former. Organisational judgement concerns the conditions under which the former contributes to the latter.

\section{Research agenda}

The argument generates empirical questions rather than formal propositions. The first concerns the Jevons hypothesis itself. Researchers should measure whether reductions in total machine-evaluation cost increase evaluation consumption, distinguishing inference cost from integration, validation, review and correction costs. Relevant outcomes include the number of evaluated cases, criteria per case, frequency of checking, location of checks in a workflow and use of repeated or recursive evaluation.

Second, research should examine substitution and complementarity. When does additional machine evaluation reduce human review time, and when does it generate more exceptions or disagreements requiring judgement? Field studies could compare workflows before and after machine evaluation becomes cheap enough for universal rather than sampled checking. The relevant dependent variable is not simply labour saved but the distribution of evaluative and authorising work across roles.

Third, evaluation abundance creates a tractable independence problem. Studies should distinguish nominal evaluator count from effective independent evidence. Kohli's (2026) findings provide a technical starting point, but organisational research should examine dependence arising from shared evidence, criteria, prompts, models and organisational assumptions. An apparently diverse panel can remain epistemically narrow if every evaluator receives the same incomplete representation.

Fourth, evaluative framing deserves direct study. Experiments can compare human judgement formed before versus after exposure to machine scores, rankings or critiques, while varying the validity of the underlying criterion. This would test whether pre-evaluation improves calibration, creates anchoring, or changes which grounds decision makers consider relevant.

Fifth, the relationship between confidence thresholds and decision rights should be examined as organisation design. Jev-like systems make it possible to route cases automatically according to estimated confidence. Research should ask who sets these thresholds, how they reflect risk tolerance, what cases become invisible to human review, and whether formal accountability remains aligned with practical control.

Finally, the connection to Executive Judgement suggests studying qualification preservation across evaluation-rich workflows. Researchers could trace successive representations of a recommendation and code whether consequential assumptions, uncertainty, provenance, dependencies and dissent survive. The decisive question is whether additional evaluation reduces qualification attrition and improves warranted commitment, or merely adds more assessments downstream of the same degraded evidence.

\section{Discussion}

The destructive comparison undertaken for this article narrows the contribution. There is insufficient basis for a new general theory of evaluation abundance. Rebound economics already explains why lower unit costs can increase consumption. Organisation theory already explains the allocation of decisions, management by exception, information overload and the difference between formal and effective authority. Automation research already explains substitution, augmentation and reliance. Ensemble and machine-evaluation research already explains why multiple outputs need not constitute independent evidence. Executive Judgement already explains why locally competent contributions can fail to become adequately warranted organisational commitment. A paper that relabelled these mechanisms would add terminology rather than explanation.

A useful residual nevertheless remains. Jev makes a specific scarcity transition technologically concrete. The relevant economic object is not intelligence in general but a bounded evaluative act embedded in a workflow. If the total marginal cost of such acts falls sufficiently, organisations may move from sampling to coverage, from periodic to persistent checking, from one criterion to many, and from final-stage review to evaluation at intermediate states. None of those design changes requires a new causal mechanism. Their scholarly importance lies in changing the feasible scale and placement at which established mechanisms operate.

This framing also resolves an apparent disagreement in the contemporary literature. Cheap-prediction accounts can be correct that judgement becomes a relatively valuable complement; post-scarcity accounts can be correct that ranking, selecting and certifying become cheap; and analytical-abundance accounts can be correct that commitment becomes harder. These claims refer to different functions. Prediction, machine evaluation, organisational judgement and authorisation need not share a cost curve. Treating them as a single category called judgement obscures rather than resolves the organisational question.

The central implication is simple: evaluation quantity is not a proxy for organisational warrant. A hundred evaluations may add little independent information if they share evidence and error structure. A perfectly consistent evaluator can scale a bad criterion. An accurate downstream check cannot recover a consequential qualification that has already disappeared from the representation being checked. A confidence threshold can change which cases receive human attention and thereby alter practical decision rights without changing the formal organisation chart. These are established mechanisms operating under a potentially different economic regime.

The paper also avoids a humanist default. Evaluation abundance does not imply that executives must personally absorb more review, nor that human judgement is intrinsically superior. Routine human evaluation may disappear; specialist governance may absorb exceptions; automated systems may themselves set or adapt some criteria; and human reviewers can be biased, overloaded or politically motivated. The empirical question is where judgement is substituted, where it is redistributed, and where new review or authorisation demands arise after the allocation of tasks and rights is taken into account.

This also clarifies the relationship with Executive Judgement. The predecessor concerns the conditions under which distributed contributions become warranted organisational commitment. The present perspective asks how those conditions behave when one class of contribution, machine evaluation, becomes much cheaper and potentially much more numerous. The two papers address different questions within the same research programme. Its value will ultimately depend on empirical work that separates evaluation count from evidence access, dependence, criterion quality, review capacity and intervention rights.

\section{Conclusion}

Jev instigates a useful organisational question because it is explicitly designed around cheap, structured machine decisions and explicitly motivated by Jevons' rebound logic. The technology does not establish that evaluation abundance has arrived, and its vendor-reported performance does not prove an organisational Jevons effect. It does, however, make a plausible transition concrete enough to investigate: bounded evaluative acts may become cheap enough that organisations consume substantially more of them.

The scholarly contribution is a focused synthesis. The article distinguishes machine evaluation from prediction, organisational judgement and authorisation; formulates a conditional Jevons hypothesis for the organisational consumption of evaluation; and positions that hypothesis against existing accounts of production abundance, post-scarcity judgement, accountable judgement and analytical abundance. Cheaper evaluation can change the feasible density and placement of checking without making the grounds for consequential commitment correspondingly abundant.

The empirical agenda follows directly. Research should establish whether lower total evaluation cost actually produces rebound, identify where additional machine evaluation substitutes for human evaluative work, and test where it instead increases exception handling, review, criterion governance or authorisation demands. It should measure effective independent evidence rather than evaluator count, distinguish formal from practical decision rights, and trace whether consequential qualifications survive into the records on which evaluation and authorisation operate.

The question raised by Jev is not whether machines can judge in the abstract. It is more operational and more organisational: what happens when evaluation becomes cheap enough to use almost everywhere? The answer cannot be inferred from evaluation volume alone. It depends on what is evaluated, against which criteria, from which evidence, with what dependence, by whom or what, and how those evaluations connect to consequential organisational commitment.

\clearpage
\section*{References}
\begingroup
\setlength{\parindent}{0pt}
\setlength{\parskip}{0.65em}

Agrawal, A., Gans, J. S. and Goldfarb, A. (2018). Human judgment and AI pricing. AEA Papers and Proceedings, 108, 58--63. \url{https://doi.org/10.1257/pandp.20181022}

Agrawal, A., Gans, J. S. and Goldfarb, A. (2019). Artificial intelligence: The ambiguous labor market impact of automating prediction. Journal of Economic Perspectives, 33(2), 31--50. \url{https://doi.org/10.1257/jep.33.2.31}

Almeida, D. (2026). Introducing System One Models \& Jev. TypeSafe AI, 15 September 2026. \url{https://typesafe.ai/blog/introducing-system-one-models-and-jev}

Brynjolfsson, E., Li, D. and Raymond, L. (2025). Generative AI at work. The Quarterly Journal of Economics, 140(2), 889--942. \url{https://doi.org/10.1093/qje/qjae044}

Cao, H., Zhu, R., Shen, H. and Xiao, Q. (2026). From production to discernment: Evaluation-centered organization design under AI-enabled production abundance. Cornell SC Johnson College of Business Research Paper, SSRN. \url{https://doi.org/10.2139/ssrn.7162118}

Addis, V. F., Attanasi, G., Di Bartolomeo, G., Mariella, M. and Peruzzi, V. (2026). Evaluating creative work with artificial intelligence: Evidence from constrained innovation tasks. Technovation, 155, 103571. \url{https://doi.org/10.1016/j.technovation.2026.103571}

Aghion, P. and Tirole, J. (1997). Formal and real authority in organizations. Journal of Political Economy, 105(1), 1--29. \url{https://doi.org/10.1086/262063}

Dietvorst, B. J., Simmons, J. P. and Massey, C. (2015). Algorithm aversion: People erroneously avoid algorithms after seeing them err. Journal of Experimental Psychology: General, 144(1), 114--126. \url{https://doi.org/10.1037/xge0000033}

Elish, M. C. (2019). Moral crumple zones: Cautionary tales in human--robot interaction. Engaging Science, Technology, and Society, 5, 40--60. \url{https://doi.org/10.17351/ests2019.260}

Hill, R. (2026). Executive Judgement in AI-Mediated Decision-Making Environments: A Process Theory of Formation, Qualification Attrition and Authorisation. arXiv preprint. \url{https://doi.org/10.48550/arXiv.2609.39442}

Hutchins, E. (1995). Cognition in the wild. MIT Press. \url{https://mitpress.mit.edu/9780262581462/cognition-in-the-wild/}

Kim, Y., Kim, J., Kim, T. and Cho, H.-C. (2026). Administrative decision-making with generative AI: The challenge of epistemic boundedness. Administration \& Society, 58(2), 284--304. \url{https://doi.org/10.1177/00953997251409156}

Kohli, G. (2026). Nine judges, two effective votes: Correlated errors undermine LLM evaluation panels. arXiv preprint arXiv:2605.29800. \url{https://arxiv.org/abs/2605.29800}

Lee, J. D. and See, K. A. (2004). Trust in automation: Designing for appropriate reliance. Human Factors, 46(1), 50--80. \url{https://doi.org/10.1518/hfes.46.1.50_30392}

Li, D., Jiang, B., Huang, L., Beigi, A., Zhao, C., Tan, Z., Bhattacharjee, A., Jiang, Y., Chen, C., Wu, T., Shu, K., Cheng, L. and Liu, H. (2025). From generation to judgment: Opportunities and challenges of LLM-as-a-judge. Proceedings of the 2025 Conference on Empirical Methods in Natural Language Processing, 2757--2791. \url{https://doi.org/10.18653/v1/2025.emnlp-main.138}

Lindebaum, D., Balasubramanian, N., Ashraf, M. and Haack, P. (2026). A process model of managerial phronesis in the age of generative AI. Academy of Management Review. Advance online publication. \url{https://doi.org/10.5465/amr.2024.0582}

Logg, J. M., Minson, J. A. and Moore, D. A. (2019). Algorithm appreciation: People prefer algorithmic to human judgment. Organizational Behavior and Human Decision Processes, 151, 90--103. \url{https://doi.org/10.1016/j.obhdp.2018.12.005}

Nonaka, I. and Toyama, R. (2007). Strategic management as distributed practical wisdom (phronesis). Industrial and Corporate Change, 16(3), 371--394. \url{https://doi.org/10.1093/icc/dtm014}

Noy, S. and Zhang, W. (2023). Experimental evidence on the productivity effects of generative artificial intelligence. Science, 381(6654), 187--192. \url{https://doi.org/10.1126/science.adh2586}

Parasuraman, R. and Riley, V. (1997). Humans and automation: Use, misuse, disuse, abuse. Human Factors, 39(2), 230--253. \url{https://doi.org/10.1518/001872097778543886}

Parasuraman, R., Sheridan, T. B. and Wickens, C. D. (2000). A model for types and levels of human interaction with automation. IEEE Transactions on Systems, Man, and Cybernetics, Part A, 30(3), 286--297. \url{https://doi.org/10.1109/3468.844354}

Raisch, S. and Fomina, K. (2025). Combining human and artificial intelligence: Hybrid problem-solving in organizations. Academy of Management Review, 50(2), 441--464. \url{https://doi.org/10.5465/amr.2021.0421}

Raisch, S. and Krakowski, S. (2021). Artificial intelligence and management: The automation--augmentation paradox. Academy of Management Review, 46(1), 192--210. \url{https://doi.org/10.5465/amr.2018.0072}

Schaefer, S. and Wickert, C. (2015). The efficiency paradox in organization and management theory. Academy of Management Proceedings, 2015(1), 10958. \url{https://doi.org/10.5465/ambpp.2015.10958abstract}

Shrestha, Y. R., Ben-Menahem, S. M. and von Krogh, G. (2019). Organizational decision-making structures in the age of artificial intelligence. California Management Review, 61(4), 66--83. \url{https://doi.org/10.1177/0008125619862257}

Shu, Y. (2026). Blind to the pivotal vote: Aggregate independence metrics miss where verification actually helps. arXiv preprint arXiv:2608.06940. \url{https://arxiv.org/abs/2608.06940}

Simon, H. A. (1955). A behavioral model of rational choice. The Quarterly Journal of Economics, 69(1), 99--118. \url{https://doi.org/10.2307/1884852}

TypeSafe AI (2026). TypeSafe AI: System One Models and Jev. Accessed 30 September 2026. \url{https://typesafe.ai/}

Baaij, M. G. (2026). Analytical abundance: Why AI makes strategy harder. California Management Review. Advance online publication. \url{https://doi.org/10.1177/00081256261470443}

Lov\'en, L. (2026). Institutions for the post-scarcity of judgment. arXiv preprint arXiv:2604.22966. \url{https://doi.org/10.48550/arXiv.2604.22966}

West, J. and Cosimano, M. (2026). Judgment Cost Theory: The accountable bearing of judgment in an age of cheap prediction. SSRN working paper. \url{https://papers.ssrn.com/sol3/papers.cfm?abstract_id=6992679}

\endgroup

\end{document}